\documentclass[aps,prl,groupedaddress,twocolumn,reprint,amsmath,amssymb]{revtex4-1}
\usepackage{graphicx}
\usepackage{dcolumn}
\usepackage{bm}
\usepackage{verbatim}

\begin{document}
\title{Quantum Storage of Photonic Entanglement in a Crystal}

\author{Christoph~Clausen}
\thanks{These authors contributed equally to this work.}
\author{Imam~Usmani}
\thanks{These authors contributed equally to this work.}
\author{F\'elix~Bussi\`eres}
\author{Nicolas~Sangouard}
\author{Mikael~Afzelius}
\author{Hugues~\surname{de~Riedmatten}}
\author{Nicolas~Gisin}

\affiliation{Group of Applied Physics, University of Geneva, CH-1211 Geneva 4,
Switzerland}

\date{\today}

\maketitle
\label{sec:abstract}
\textbf{Entanglement is the fundamental characteristic of quantum physics. Large
experimental efforts are devoted to harness entanglement between various physical
systems. In particular, entanglement between light and material systems is
interesting due to their prospective roles as ``flying'' and stationary qubits in
future quantum information technologies, such as quantum 
repeaters~\cite{Briegel1998,Duan2001,Sangouard2009} and quantum 
networks~\cite{Kimble2008}. Here we report the first demonstration of entanglement between a photon at
telecommunication wavelength and a single collective atomic excitation stored in
a crystal. One photon from an energy-time entangled pair~\cite{Franson1998} is
mapped onto a crystal and then released into a well-defined spatial mode after a
predetermined storage time. The other photon is at telecommunication wavelength
and is sent directly through a 50~m fiber link to an analyzer. Successful
transfer of entanglement to the crystal and back is proven by a violation of the
Clauser-Horne-Shimony-Holt (CHSH) inequality~\cite{Clauser1969} by almost three
standard deviations ($\bf S=2.64 \pm 0.23$). These results represent an important
step towards quantum communication technologies based on solid-state devices. In
particular, our resources pave the way for building efficient multiplexed quantum
repeaters~\cite{Simon2007,Usmani2010} for long-distance quantum networks.}

\label{sec:introduction}
While single atoms~\cite{Blinov2004,Volz2006} and cold atomic 
gases~\cite{Matsukevich2004, Matsukevich2005, deRiedmatten2006, Chen2007a, Sherson2006,
Jin2010} are currently some of the most advanced light-matter quantum interfaces,
there is a strong motivation to control light-matter entanglement with more
practical systems, such as solid-state devices~\cite{Togan2010}. Solid-state
quantum memories for photons can be implemented with cryogenically cooled
crystals doped with rare-earth-metal (RE) ions~\cite{Tittel2010}, which have
impressive coherence properties at temperatures below 4~K. These solid-state
systems have the advantage of simple implementation since RE-doped crystals are
widely produced for solid-state lasers, and closed-cycle cryogenic coolers are
commercially available. Important progress has been made over the last years in
the context of light storage into solid-state memories, including long storage
times~\cite{Longdell2005}, high efficiency~\cite{Hedges2010} and storage of light
at the single photon level with high coherence and negligible 
noise~\cite{Riedmatten2008, Lauritzen2010,Usmani2010,Hedges2010,
Chaneliere2010,Sabooni2010}. Yet, these experiments were realized with classical
bright or weak coherent states of light. While this is sufficient to characterize
the performances of the memory, and even to infer the quantum characteristics of
the device~\cite{Riedmatten2008, Hedges2010}, it is not sufficient for the
implementation of more sophisticated experiments involving entanglement, as
required for most applications in quantum information science. For this purpose,
it is necessary to store non-classical light, in particular individual photons
that are part of an entangled state. In addition, for quantum communication
applications, the other part of the entangled state should be a photon at
telecommunication wavelength in order to minimize loss during transmission in
optical fibers.

In this Letter, we report on an experiment where a photon from an entangled pair
is stored in a quantum memory based on a RE-doped crystal. The quantum properties
of the photon persist after the retrieval from the memory. More specifically, we
show that the non-classical nature of the intensity correlations is preserved,
and that the entanglement fidelity of the retrieved photon is sufficient to
violate a Bell inequality. Such a violation is explicitly demonstrated.
These results represent the first successful demonstration of the reversible mapping of
energy time entangled photons onto and out of a quantum memory.

\section{Experiment}
\begin{figure}
\begin{center}
  \includegraphics[width=\linewidth]{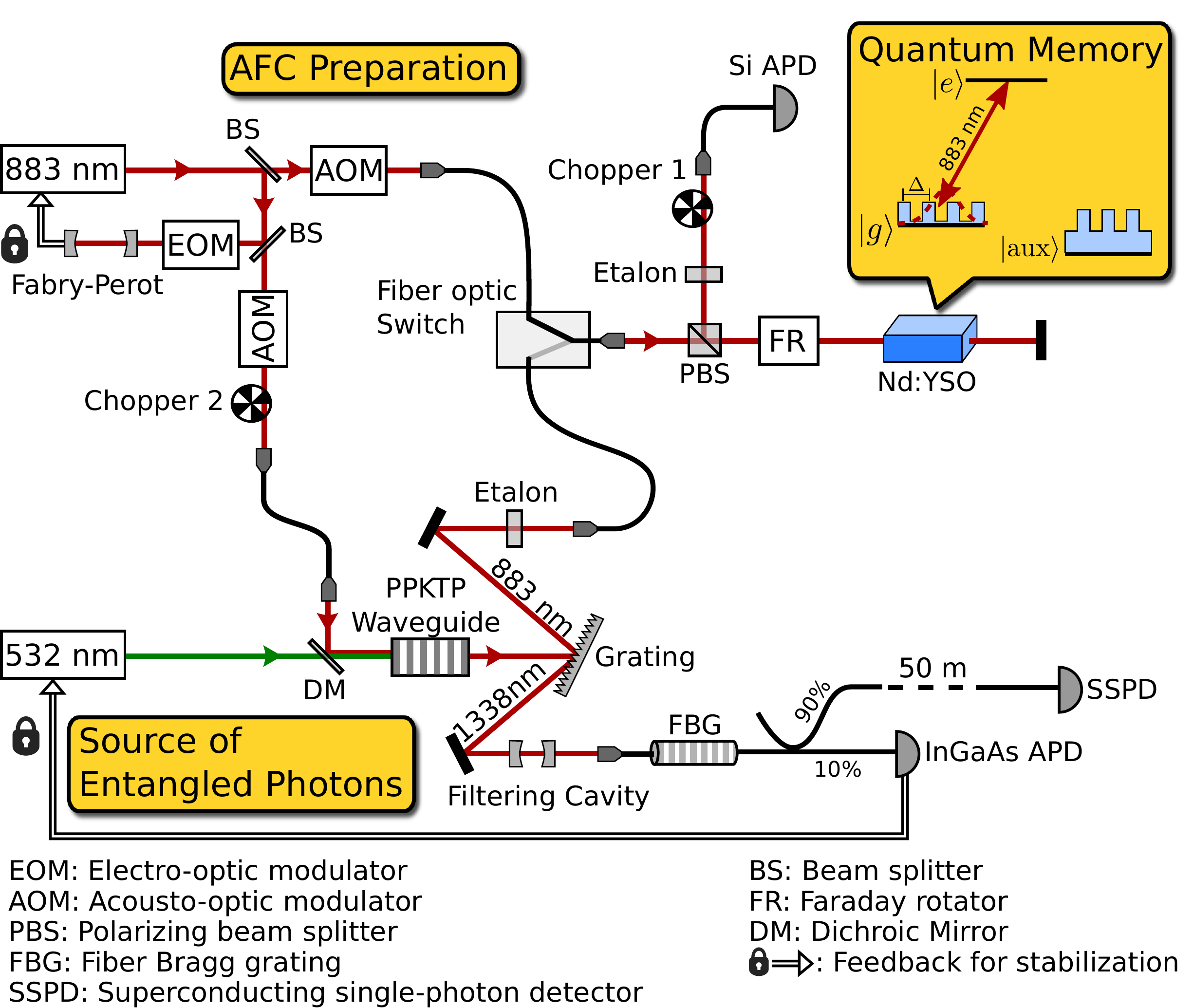}
  \caption{\textbf{Experimental setup}. The experimental setup can be divided
  into three parts: The crystal serving as quantum memory, the laser system for
  the preparation of the atomic frequency comb (AFC) in the crystal, and the
  source of entangled photons with associated spectral filtering. During the experiment we
  periodically switch between 15~ms of AFC preparation followed by a
  15~ms measurement phase, where entangled photons are stored. During the
  preparation, the comb structure is prepared by frequency-selective optical pumping using light from 
  an 883~nm diode laser in combination with an
  acousto-optic modulator. The fiber optic switch is in the upper position, and
  the silicon avalanche photodiode (Si APD) is protected from the bright light
  by chopper 1. At the same time, we must ensure that the central frequency of
  the optical filtering system at 1338 nm and of the AFC at 883 nm both satisfy
  the energy conservation of the SPDC process. To this end, chopper 2 is opened,
  and a small fraction of the light at 883~nm is overlapped with the light of the 532~nm continuous wave
  laser that pumps the PPKTP waveguide. This leads to the creation of light at
  1338~nm by difference frequency generation (DFG).
  Using this DFG signal, the frequency of the 532~nm light is
  adjusted such the detection rate on a separate InGaAs APD stays
  constant, which means that the 1338~nm DFG light is in resonance with the
  filtering cavity. Long-term stability of the 883~nm laser itself is achieved
  by continuously referencing it to a Fabry-Perot cavity.
  During the measurement phase, the positions of switch and choppers are
  reversed. Now, entangled pairs of photons are generated in the waveguide via
  SPDC. The two photons in a pair are spatially separated by a diffraction
  grating and then strongly filtered. The
  filtering at 1338~nm consists of a cavity followed by a fiber Bragg grating
  that removes spurious longitudinal modes of the cavity. Then the photons are
  directed towards a superconducting single photon detector located in another 
  laboratory 50~m away. The photons at
  883~nm undergo filtering by two etalons with different free spectral ranges,
  such that only photons within a narrow range of frequencies pass through both
  etalons. The photons are sent through the crystal in a double-pass configuration 
  to increase the absorption probability,
  and are afterwards detected by the Si APD. See Supplementary Information for
  more details.}
   \label{fig:afc_setup}
\end{center}
\end{figure}

Our experiment consists of a coherent solid-state quantum memory and a source of
entangled photons. The complete experimental setup, comprising the source of
entangled photon pairs and quantum memory, is shown in Fig.~\ref{fig:afc_setup}.

The source is based on non-degenerate spontaneous parametric down conversion
(SPDC) in a non-linear PPKTP waveguide pumped by continuous wave light at
532~nm, yielding energy-time entangled photons with the signal photon at the memory wavelength of
883~nm, and the idler photon at the telecom wavelength of 1338~nm. Both photons
initially have a spectral width of approximately 1.5~THz, a factor of $10^4$
larger than the 120~MHz bandwidth of the memory. Hence, strong filtering is
crucial~\cite{Akiba2009} to achieve signal-to-noise ratios sufficiently large to
reveal the presence of entanglement after the storage (see
Fig.~\ref{fig:afc_setup} and Supplementary Information). Additionally, active
frequency stabilization was necessary to maintain the simultaneous spectral
overlap of the photon pairs with the filtering system and the quantum memory
(see Supplementary Information).

The quantum memory is a Y$_2$SiO$_5$ crystal impurity-doped with neodymium ions
having a resonance at 883~nm with good coherence properties~\cite{Usmani2010}. It
is based on a photon echo type interaction using an atomic frequency comb 
(AFC)~\cite{Afzelius2009a}. In an AFC the absorption profile of the atomic ensemble is
shaped into a comb-like structure by optical pumping. A photon is then, with some
efficiency, absorbed and re-emitted into a well-defined spatial mode due to a
collective rephasing of the atoms in the comb structure. The time of re-emission
depends on the period of the comb and is pre-determined.
Using weak coherent states, we have previously shown that our memory is capable of
coherently storing multiple temporal modes~\cite{Usmani2010}. Therefore, this
type of interface is perfectly suited for storing true single photons created at
random times, as is the case for energy-time entanglement. Nevertheless,
switching to photons generated by SPDC posed major challenges. Besides the
elaborate filtering of the photons and its associated frequency stabilization, it
was necessary to significantly increase the storage efficiency. Employing a new
optical pumping scheme for the preparation of the AFC (see Supplementary
Information), the efficiency was increased by a factor of 3 for storage times
below 200~ns, now reaching values up to 21\% (see results below). After being
released from the memory, the photon at 883~nm is detected with a single-photon
avalanche photodiode. The filtered photon at 1338~nm is sent to another
laboratory 50~m away via a single mode fiber, where it is analyzed and detected
using a superconducting single photon detector (SSPD). Owing to the low loss at
telecommunication wavelengths,  the amount of optical fiber could in principle
be extended to several kilometres without significantly affecting the results
presented herein.

\section{Results}
\subsection{Non-classical correlations}
\begin{figure*}
\begin{center}
  \includegraphics[width=\linewidth]{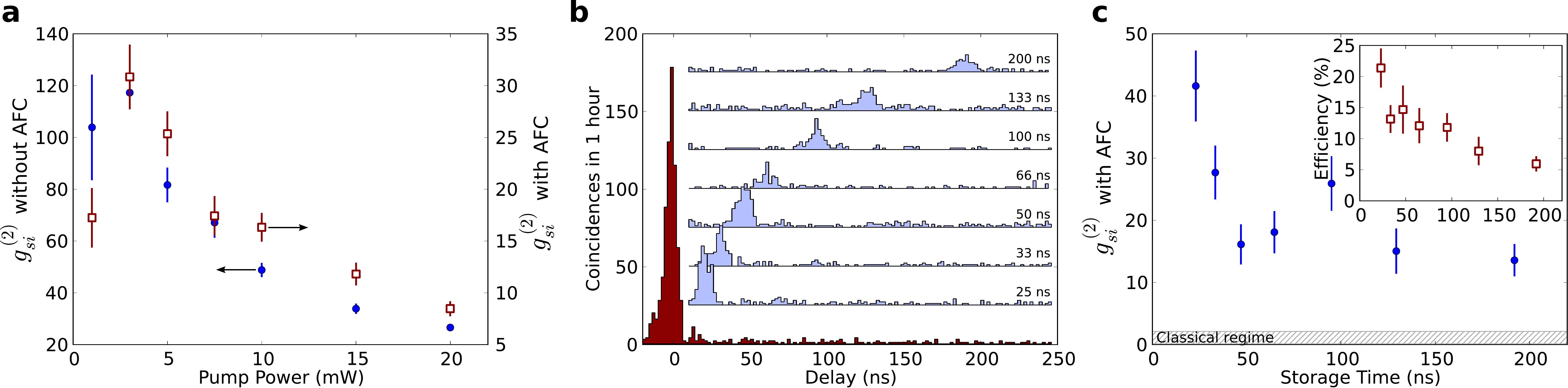}
  \caption{\textbf{Measurements of the cross-correlation and storage
  efficiency}. (a) Cross-correlation $g_{si}^{(2)}$ as a function of
  the pump power. Data points shown were taken with an
  AFC memory storage time of 25~ns (empty squares), and for comparison, with the
  crystal prepared with a 120~MHz wide transmission window, i.e.~without AFC (filled circles).
  We achieve the highest correlations for a pump power around 3~mW. The size of the
  coincidence window is 10~ns. (b) Coincidence histograms for different
  storage times. Individual histograms are vertically offset for clarity. Each
  histogram shows a peak at the pre-determined storage time. For comparison, the
  bottom-most histogram (red) was taken without AFC.
  (c) Cross-correlation $g_{si}^{(2)}$ as a function of storage
  time with 10~ns coincidence window, extracted from (b). For storage times
  up to 200~ns the correlations stay well above the classical regime
  given by Eq.~\eqref{eq:non-classicality} (shaded area). The inset shows the
  storage efficiency for the same range of storage times.}
  \label{fig:g12-histograms}
\end{center}
\end{figure*}
In a first experiment we verified that the non-classical nature of the intensity
correlations between the two photons is preserved during the storage and
retrieval process. Neglecting the exact frequency dependence, the state of the
photons created in the SPDC process, provided that the pair creation probability
$p$ is much smaller than 1, is described by
\begin{equation}
  \label{eq:two-mode-squeezed-state}
  |0_s,0_i\rangle + \sqrt{p}|1_s,1_i\rangle + O(p),
\end{equation}
where the subscript $s$ $(i)$ indicates the
signal (idler) mode at 883~nm (1338~nm). While in such a state both the signal
and idler modes individually exhibit the statistics of a classical thermal
field, its quantum nature can be revealed by
strong intensity correlations between the two modes. Indeed, assuming 
second-order auto-correlations of signal and idler of
$g_{s}^{(2)} = g_{i}^{(2)} = 2$, a lower bound of the cross-correlation function
$g_{si}^{(2)}$ gives a criterion for non-classicality~\cite{Lee1990},
\begin{equation}
  \label{eq:non-classicality}
  g_{si}^{(2)} = \frac{P_{si}}{P_s P_i} > \frac{1}{2}\left( g_s^{(2)} + g_i^{(2)}\right) = 2,
\end{equation}
where $P_{s}$ ($P_{i}$) is the probability to detect a signal (idler) photon,
and $P_{si}$ the probability for a coincidence detection. In practice, $P_{si}$
($P_s P_i$) is determined by the number of coincidences in a certain time window
centered on (away from) the coincidence peak.
 
To find the optimum conditions for the experiment, we first measured the
cross-correlation as a function of the pump power of the source, as
shown in Fig.~\ref{fig:g12-histograms}a. In one measurement series we used an AFC
with 25~ns storage time. For comparison, we took a second series with the crystal
prepared without the AFC, that is, with a 120~MHz wide transmission window.
In both cases, the results are typical for photons from an SPDC source. For low pump powers, the cross-correlation is limited by detector dark counts, and at high pump powers
it is reduced by the contribution of multiple pairs, i.e.~the
higher order terms in \eqref{eq:two-mode-squeezed-state}. We find an optimum around a
pump power of 3~mW, where $g_{si}^{(2)} \simeq 115$ without AFC, and
$g_{si}^{(2)} \simeq 30$ after a 25~ns storage, thus proving the quantum
character of the storage. The reduction in the cross-correlation with the
storage is due to limited efficiency ($21$\%). This can be
considered as loss which effectively increases the contribution of accidental
coincidences stemming from dark counts and multiple pair emissions.

Next, we measured the memory efficiency and the cross-correlation for different
storage times, as shown in Fig.~\ref{fig:g12-histograms}b and
\ref{fig:g12-histograms}c. The efficiency is defined as the ratio of the number
of stored and released photons to the total number of photons incident on the
crystal, and includes effects of decoherence in the memory~\cite{Usmani2010}. The
longer the storage time, the more difficult it is to prepare a comb with optimal
shape (see Supplementary Information). This is due to material limitations and
leads to a decrease of efficiency, and consequently to a decrease of the
cross-correlation. However, the latter stays well above the classical limit for
storage times up to 200~ns. Possibilities to extend this storage time will be discussed
below.

\subsection{Entanglement}
We now turn our attention towards the most particular kind of
quantum correlations, namely entanglement. By performing a two-photon quantum
interference experiment, we show that the entanglement of the photon pair is
preserved when the signal photon is stored in the crystal.

\begin{figure*}
\begin{center}
  \includegraphics[width=\linewidth]{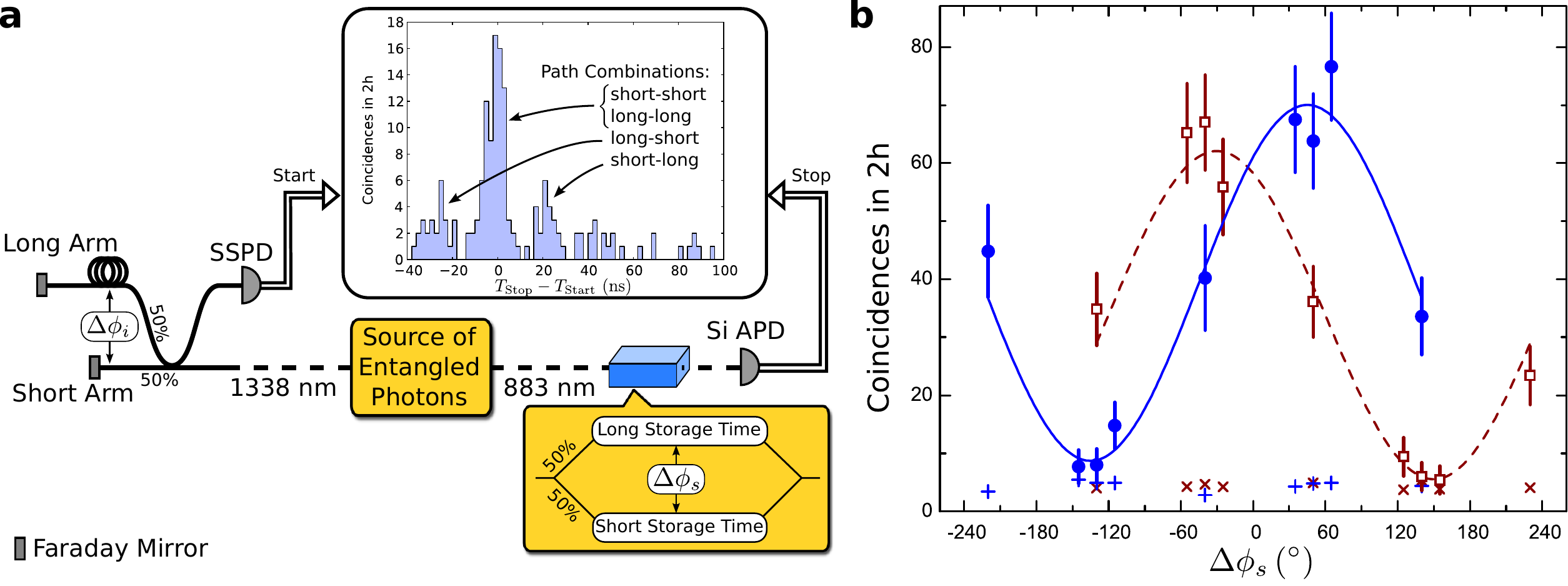}
  \caption{
  \label{fig:interference}
  \textbf{Entanglement preservation during the quantum storage}.
  (a) To reveal the presence of energy-time entanglement, we used a Franson-type
  setup. A fiber interferometer with $\tau=25$~ns path length difference was
  inserted before the detector of the idler photon. For the signal photon, the
  interferometer was implemented using the quantum memory with partial
  read-outs with a short storage time of 50~ns and a long storage time of 75~ns.
  The resulting coincidence histogram shows three peaks separated by $\tau$ corresponding to different path combinations. Due to interference, the coincidence probability at zero time-delay, i.e.~in the central peak, oscillates as a
  function of both of the relative phases $\Delta \phi_{s}$ and $\Delta
  \phi_{i}$, given by Eq.~\eqref{eq:interference-rate}. (b) Number of
  coincidences in the central peak in two hours as a function of the relative phase $\Delta \phi_s$ for two
  values of $\Delta \phi_i$. The pump power was 5~mW, and the size of the
  coincidence window 10~ns. The solid and dashed lines result from fits to 
  Eq.~(\ref{eq:interference-rate}) and respectively give visibilities of $V=78
  \pm 4 \%$ and $84 \pm 4\%$. The visibilities are mainly limited by the level
  of accidental coincidences (crosses). The fit also gives a difference between 
  the two values of $\Delta \phi_i$ of $75\pm10^\circ$. These values closely
  match settings necessary for a maximal violation of the CHSH inequality.}
\end{center}
\end{figure*}

Photon pairs generated by our source are energy-time entangled, that is, the two
photons in a pair are created simultaneously to ensure energy conservation, but
the pair creation time is uncertain to within the coherence time of the pump
laser. We want to reveal the presence of this entanglement using a Franson-type
setup~\cite{Franson1998}, see Fig.~\ref{fig:interference}a. Such a setup requires
two equally unbalanced interferometers with delay $\tau$ larger
than the coherence time of the photons (here about 5~ns), so there
can be no single-photon interference. If additionally the delay is
shorter than the coherence time of the pump, the uncertainty in the creation
time will lead to quantum interference between the two path combinations short-short and
long-long. The interference manifests itself in the coincidence probability~\cite{Franson1998},
\begin{equation}
  \label{eq:interference-rate}
  P_{si} \propto 1 + V\cos(\Delta \phi_s + \Delta \phi_i),
\end{equation}
where $V$ is the visibility of interference, and $\Delta \phi_{s,i}$ are the
relative phases acquired between the short and long arms in the signal and idler
interferometers, respectively. The observed interference can also be interpreted
as a post-selection of a time-bin entangled state,
\begin{equation}
  \label{eq:energy-time}
  \frac{1}{\sqrt{2}}\left(|E_s E_i\rangle + |L_s
  L_i\rangle\right),
\end{equation}
with a time difference $\tau=25$~ns between the early ($E$) and late ($L$)
time-bins. For the idler photon we used a fiber interferometer,
while for the signal photon, we implemented the interferometer by the technique
of partial read-outs~\cite{Riedmatten2008} of the AFC at two storage times
separated by $\tau$, which gives us excellent control over $\Delta \phi_s$. Note
that the temporal multi-mode
capability~\cite{Usmani2010,Afzelius2009a,Nunn2008} is necessary to coherently
map the signal time-bin qubit onto a collective atomic excitation upon absorption.

Figure~\ref{fig:interference}b shows the measured coincidence rate as a function
of $\Delta \phi_s$ for two values of $\Delta \phi_i$ and a source pump power of
5~mW. The visibility of the interference is $V=84 \pm 4\%$ and $78\pm 4\%$,
which corresponds to a mean two-qubit fidelity with the input state of $F=86 \pm 2\%$. Hence, according to the Peres
criterion~\cite{Peres1996} of $F\geq 0.5$, the two photons are still entangled
after the quantum storage. This also means that while the signal photon was
still inside the crystal, the post-selected entangled state (neglecting noise)
was of the form
\begin{equation}
  \frac{1}{\sqrt{2}}\left(|E_{QM} E_i\rangle + |L_{QM}
  L_i\rangle\right),
\end{equation}
that is, the idler photon was entangled with a collective atomic excitation
inside the quantum memory, denoted by the subscript $QM$.

The nonlocal character of the entanglement can be revealed by a violation of the
Clauser-Horne-Shimony-Holt (CHSH) inequality~\cite{Clauser1969}. Note that the
possibility to violate this inequality can be inferred indirectly from a
visibility larger than 71\%. Nevertheless, we performed the measurements
necessary for a direct violation of the inequality and obtained $S=2.64\pm0.23$
(details are given in the Supplementary Information). This proves the presence of nonlocal
quantum correlations between the telecom photon and a collective atomic
excitation in the crystal, and hence the presence of entanglement independent of
the experimental details~\cite{Acin2006a}. At the same time, this would guarantee
the security against individual attacks for quantum key distribution applications~\cite{Acin2006a}.

\subsection{Bell test involving hybrid qubit}
A particularly intriguing situation arises when instead of using two modes of
the partial read-out as a time-bin qubit, one uses a single read-out together with the
transmitted part of the photon as time-bins. Indeed, the imbalance between the storage efficiency and the
transmission probability offers a well suited qubit analyzer for a violation 
of the CHSH inequality using bases lying in the $xz$-plane of the Bloch sphere~\cite{Bussieres2010} (see Supplementary Information). We performed such a
measurement and observed $S=2.62\pm 0.15$, i.e.~a violation by more than 4 standard deviations. This is
an entanglement witness showing that
at any step before the detections, the initial entanglement
\eqref{eq:energy-time} was preserved. In particular, this implies that while the
early mode was stored in the crystal, the two-qubit state
(neglecting noise and omitting normalisation) was of the form
\begin{equation}
  \label{eq:hybrid}
  \alpha|E_{QM} E_i\rangle + |L_s L_i\rangle,
\end{equation}
where $\alpha$ is related to the absorption efficiency $\eta_{\rm abs}$
of the quantum memory by $\alpha = \sqrt{\eta_{\rm{abs}}}$.
This is an entangled state between a telecommunication-wavelength qubit and a
photon-crystal hybrid qubit. We note that this kind of hybrid qubit is
the key ingredient of one of the most efficient quantum repeaters based on atomic
ensembles and linear optics~\cite{Sangouard2008}.

\section{Outlook}
The fact that one can create entanglement between a single photon and a
macroscopic object -- in this case a collective atomic excitation delocalized
over a 1~cm long crystal -- is fascinating in itself. Beyond
their fundamental interest, the results presented in this Letter are part of the effort
towards the long-term goal of the implementation of an efficient quantum
repeater. Quantum repeaters are a promising solution to the problem of finite
loss in optical fibers, which at the moment limits the practical distance for
entanglement distribution and quantum cryptography~\cite{Sangouard2009}. Our
results show that commercial crystals can be used as quantum memory for entangled photons. This
represents an important enabling step towards solid state based quantum
repeaters. The next major challenges are clearly identified: longer storage times, the possibility
for on-demand read-out of the memory and higher efficiency. Promising ideas and
encouraging results in these individual directions do already exist. Crystals of
Y$_2$SiO$_5$ doped with praseodymium or europium have a suitable energy-level
structure and very good coherence properties. Indeed, on-demand read-out has
been demonstrated with bright pulses in a praseodymium-doped 
crystal~\cite{Afzelius2010}, and europium adds the potential of larger bandwidth, higher
multimode capacity and long storage times~\cite{Afzelius2009a}. The efficiency is
directly linked to the optical depth of the material~\cite{Afzelius2009a}, and
can, for example, be addressed by using longer crystals~\cite{Hedges2010}.
Alternatively, optical cavities can be used to increase the effective interaction
length~\cite{Afzelius2010a,Moiseev2010} without material specific side effects.
It will be exciting to follow how these developments lead to
the actual realisation of a quantum repeater.

We note that, parallel to this work, Saglamyurek \emph{et~al.} have also demonstrated 
storage and retrieval of an entangled photon using a thulium-doped lithium niobate waveguide~\cite{Saglamyurek2010}.

\textbf{Acknowledgements.}  We thank Reto Locher for help during the early
stages of the experiment. We are very grateful to Alexios Beveratos and Wolfgang Tittel
for lending us APDs. This work was supported by the Swiss NCCR Quantum
Photonics, as well as by the European projects QuRep and ERC-Qore. F.B. was
supported in part by le Fonds qu\'eb\'ecois de la recherche sur la nature et les
technologies (FQRNT).

\section{Supplementary Information}
\subsection{Atomic Frequency Comb}
The atomic frequency comb (AFC) memory is a photon-echo based scheme, where the
absorption profile of the crystal is shaped into a comb-like structure using
optical pumping, thus making it possible to take full advantage of the high
atomic density in a doped crystal, despite of inhomogeneous broadening~\cite{Afzelius2009a}.

When a photon enters the crystal, with a spectral bandwidth covering a large part
of the AFC spectrum, it is absorbed, provided that the optical depth is
sufficient. After the absorption, the photon is stored in a single atomic
excitation delocalized over all the atoms, corresponding to a collective Dicke type state,
\begin{equation}
  |\Psi(t)\rangle \propto \sum_j^N c_j e^{-ikz_j}e^{i 2 \pi\delta_j
  t} |g_1 \cdots e_j \cdots g_N\rangle.
\label{dicke}
\end{equation}
In the AFC the distribution of atomic detunings $\delta_j$ is periodic with
period $\Delta$. After a time $t_s=1/\Delta$, the components of the state
(\ref{dicke}) are in phase and lead to the reemission of a photon in a well
defined spatial mode $\vec{k}$ with high probability. 

\begin{figure}
\begin{center}
  \includegraphics[width=\linewidth]{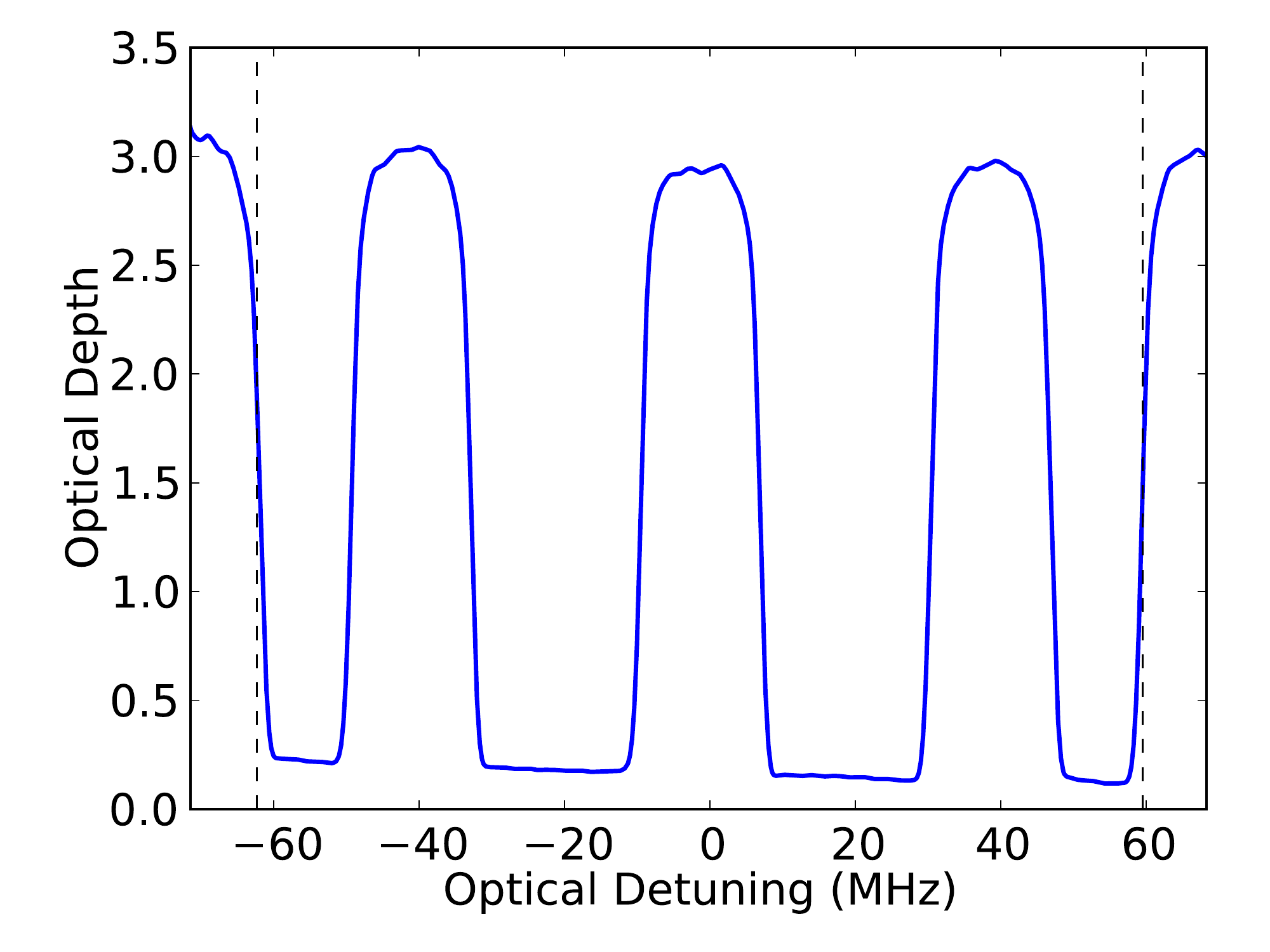}
  \caption{\textbf{Spectrum of an
  AFC prepared by an incoherent optical pumping technique.} The peaks have
  a width larger than the effective linewidth of the material, so it is
  possible to use the whole available optical depth and to have
  almost square-shaped peaks, which optimizes the rephasing. The dashed lines
  delimit the 120~MHz bandwidth of the AFC.}
\label{fig:comb}
\end{center}
\end{figure}

Here we used a storage medium composed of $ \sim 10^9$ $\text{Nd}^{3+}$ dopant
ions in an $\text{Y}_2\text{SiO}_5$ crystal cooled to 3~K. This crystal has
already shown a high multimode capacity~\cite{Usmani2010} with efficiencies
between 6\% (for $t_s=100$~ns) and 1\% ($t_s=1.5$~$\mu$s). In the work
presented here we have improved the storage efficiency in the range $t_s=25$ to
200 ns by creating comb peaks with an approximate square shape. This has been shown to be the optimal
shape in terms of storage efficiency~\cite{Chaneliere2010}. Note also that the
AFC structure is prepared by using an incoherent optical pumping technique
consisting of simultaneously sweeping the laser frequency and modulating its
intensity, as also used in~\cite{Lauritzen2010}. As a result we observe an
efficiency of 21\% for $t_s=25$ ns and 12\% for $t_s=100$ ns. The decay of the
efficiency with longer storage time (thus for closer comb spacing $\Delta$) is
due to limitations in the coherence properties of the material, which
deteriorates the comb shape for closely spaced peaks. We refer to Ref.~\cite{Usmani2010}
for a more detailed discussion on this optical pumping
issue. For the shortest storage time of 25 ns, however, this problem was almost
negligible. Hence, the obtained efficiency is mainly limited by the
absorption depth of the crystal. An example of a close to optimal comb for
$t_s=25$ ns is shown in Fig.~\ref{fig:comb}. Using the measured comb as an input to a numerical
Maxwell-Bloch simulation gave an efficiency close to the measured one.

\subsection{Spectral filtering of photon pairs}
\begin{table*}
	\caption{\textbf{Spectral filtering of photon pairs.}Overview of bandwidth and
	efficiency for the elements of the optical setup, where the efficiency for optical elements equals the peak
	transferred intensity normalized to incoming intensity. The free spectral
	ranges (FSR) of etalons and cavity, as well as the dark count rate (DC) of the
	detectors, are given in parenthesis.}
	\label{tbl:losses}
	\begin{tabular}{cp{6cm}cr}
	Wavelength & Element & Bandwidth & Efficiency\\ \hline
	\hline
	883~nm & Grating & 90~GHz & 70\%\\
	 & Solid etalon (FSR 42~GHz)& 600~MHz  & 80\%\\
	 & Air-spaced etalon (FSR 50~GHz)& 600~MHz  & 80\%\\
	 & AFC & 120~MHz & --\\
	 & Fiber coupling, polarization controller, fiber optic switch, mirrors, lenses
	 and windows & -- & 4\%\\ & Detector (DC 100~Hz) & -- & 30\%\\ \hline
	 & Total & 120~MHz & 0.5\%\\ \hline \hline
	1338~nm & Grating & 60~GHz & 90\%\\
	 & Filter Cavity (FSR 24~GHz) & 45~MHz & 30\%\\
	 & Fiber Bragg Grating & 16~GHz & 50\%\\
	 & Fiber coupling, circulator, polarization controller, fiber beam splitter and
	 mirrors & -- & 14\%\\ & Detector (DC 10~Hz) & -- & 8\%\\ \hline & Total &
	 45~MHz & 0.15\%\\\hline\hline
	\end{tabular}
\end{table*}

An overview of the various filtering steps for each of the two wavelengths is
given in Table~\ref{tbl:losses} (see also Fig.~\ref{fig:afc_setup}). As a first
filtering step a diffraction grating spatially separates the pump, signal and idler photons and,
in combination with coupling into single-mode fibers, it reduces the bandwidth
of the photons at 883 (1338)~nm to 90 (60)~GHz. At 1338~nm, the
photons are filtered to their final linewidth of 45~MHz by coupling through a
Fabry-Perot cavity with free spectral range (FSR) 23.9~GHz. Additional
longitudinal modes of the cavity are subsequently removed by a Fiber Bragg
grating with 16~GHz bandwidth.

Filtering one of the photons in the pair is the same as filtering the photon pair
as a whole, since energy conservation guarantees that photons measured in
coincidence have the same bandwidth. However, uncorrelated photons would then
contribute significantly to the accidental coincidences. Therefore, additional
filtering at 883~nm was necessary. To this end, we use one solid
and one air-spaced etalon, both with bandwidths around 600~MHz. Different FSRs
of 42 and 50~GHz eliminate spurious longitudinal modes. Additionally, outside the
120~MHz bandwidth of the AFC the absorption of the crystal with an inhomogeneous
linewidth of 6~GHz provides a final filtering step.

We use detectors with 30\% detection efficiency and 100~Hz dark counts at
883~nm, and 8\% and 10~Hz at 1338~nm. The overall detection
efficiencies, which are detailed in Table~\ref{tbl:losses}, are
approximately equal to 0.15\% at 1338~nm and 0.5\% at 883~nm. This includes
losses of the switch and all other optical elements.

\subsection{Measurements and frequency stabilization}
All experiments consisted of a two-part cycle
with 15~ms used for the preparation of the AFC and frequency stabilization of the
setup, and 15~ms of actual measurement where photons are stored. A fiber optic
switch allows for rapid switching between preparation and measurement, and a combination of
choppers and acousto-optic modulators efficiently protects the highly
sensitive Si APD from the preparation light, which at the position of the
crystal has a power of about 1~mW.

The experimentally relevant quantity in all our measurements is
coincidence statistics, i.e.~the time between the detection of the
telecom photon and its partner at 883~nm. Typical rates are a
few coincidences per minute. With accumulation times reaching several hours, a
high degree of frequency stability of the lasers and filtering elements are
indispensible. In particular drifts of the AFC preparation
laser with respect to the pump laser of the source had to be eliminated.
Otherwise, the photon-pair frequencies $\omega_{883} + \omega_{1338} =
\omega_{532}$ imposed by energy conservation in the SPDC would not simultaneously match the center of
the AFC and that of the filtering system at 1338~nm. Drifts were eliminated
using the following method. First, the long-term stability of the 883~nm laser
was dramatically increased by locking it to a temperature stabilized Fabry-Perot
cavity. Second, during the 15~ms preparation cycle, we injected a fraction of
the 883~nm light into the waveguide. The frequency of the light created at 1338~nm via
difference frequency generation (DFG) was then locked to the filtering cavity
using a side-of-fringe technique. As a result, long-term frequency deviations between the
center of the AFC structure and the filtered photon pairs were reduced to about
1~MHz over several hours.

For measurements involving the unbalanced interferometer for the telecom photon,
the phase of the interferometer was also stabilized using the highly
coherent DFG light.

\subsection{CHSH inequality with partial read-outs}
We give here some experimental details of the violation of the CHSH
inequality~\cite{Clauser1969} using the entangled photon pairs after the signal
mode has been stored in and released from the AFC quantum memory.

Our source produces photon pairs entangled in energy and time~\cite{Franson1998}.
Due to post-selection we can simplify the following treatment by considering just
two particular temporal modes delayed by $\tau$, the early and late modes
\begin{equation}
  \label{ini}
  \frac{1}{\sqrt{2}}\left(|E_sE_i\rangle+|L_sL_i\rangle\right),
\end{equation}
where $\tau$ is larger than the coherence time of the individual modes.
The usual way to highlight time-bin entanglement is to measure two-photon
interference in a Franson setup~\cite{Franson1998} involving two unbalanced
interferometers with path length difference $\tau$. By choosing appropriately
the measurement settings, i.e.~the phases and the beam splitting ratio of the interferometers, the non-local characteristic of this interference can be revealed through the violation of a Bell inequality, e.g.~the CHSH inequality. Interestingly, Bell tests can be
seen as entanglement witnesses proving that, at any step before the detection,
the entanglement produced at the level of the source is preserved, at least 
partially; see below.

In our setup (see Fig.~\ref{fig:afc_setup}), we use a fiber interferometer for
the idler photon. For the signal photon, the interferometer is implemented using partial read-outs of the quantum
memory~\cite{Riedmatten2008}. 

\begin{figure}
\begin{center}
  \includegraphics[width=\linewidth]{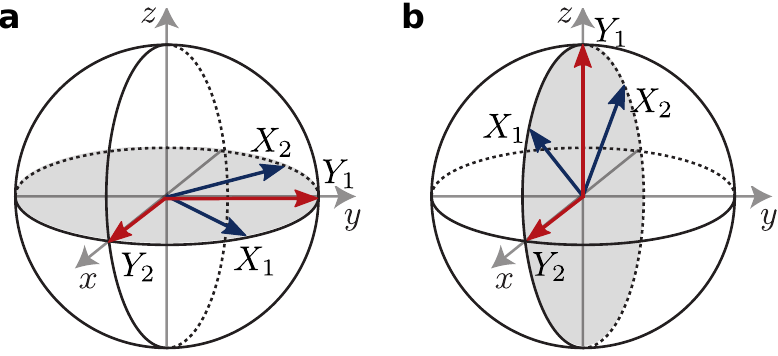}
  \caption{\textbf{Measurement bases for the violation of the CHSH inequality.}
  Blue (red) arrows indicate signal (idler) bases $X_1$ and $X_2$ ($Y_1$ and
  $Y_2$). (a) For the violation of the CHSH inequality, all four bases lie on the equator of the Bloch sphere with the appropriate angles between them. (b) The hybrid qubit represents a state will unequal
  populations, and optimal violation can only be obtained on a plane including
  the poles of the Bloch sphere.}
  \label{fig:bloch-sphere}
\end{center}
\end{figure}

The measurement of the CHSH inequality requires two bases for the signal photon
$X_1$ and $X_2$, and similarly $Y_1$ and $Y_2$ for the idler photon. To obtain
maximum violation, a possible choice of bases is shown in
Fig.~\ref{fig:bloch-sphere}a. All four bases lie on the equator of the Bloch
sphere, $X_1$ and $X_2$ form an angle of $90^\circ$, and $Y_1$ and $Y_2$ are
obtained by rotating $X_1$ and $X_2$ by $45^\circ$.
This is the case here, since the probabilities to detect a
photon that has taken a long or short path is the same, for both signal and
idler; see Fig.~\ref{fig:interference}a. The angles between the bases are given
by the relative phases induced by the fiber interferometer and the storage in the AFC. The relative phase in the AFC is
controlled by shifting the comb pattern with respect to the central frequency of
the signal photon~\cite{Afzelius2009a}. In the fiber interferometer, we use the light
produced by DFG to set the phase. This light has a coherence time much longer
than $\tau=25$~ns and does show single-photon interference in the
interferometer. The interferometer is then stabilized to maintain a certain point on the resulting interference fringe.

Since we had only one detector in each side, we had to add
a $\pi$-phase shift every time we wanted to access the second outcome of the
chosen basis. For this reason, we had to make 4 measurements per correlator, i.e.~16
for the whole Bell test. For a pump power of 3~mW we obtained
\begin{eqnarray}
  \nonumber && E(X_1 Y_1)=0.68\pm 0.12,\\
  \nonumber && E(X_2 Y_1)=0.79\pm 0.10,\\
  \nonumber && E(X_1 Y_2)=0.60\pm 0.10,\\
  \nonumber && E(X_2 Y_2)=-0.57\pm 0.14,
\end{eqnarray}
leading to $S=2.64\pm 0.23$. Here, the uncertainties are standard deviations
related to the poissonian statistics of the coincidence events. This is a clear
violation of the CHSH inequality and proves not only the conservation of the
entanglement during the storage, but that the telecom photon was entangled with
a collective excitation in the crystal.

\subsection{CHSH inequality with a hybrid qubit}
The violation of the CHSH inequality using the hybrid qubit is
done in analogue to the violation with partial read-outs described above.
However, we now want to reveal the entanglement between a photonic time-bin qubit, at telecommunication wavelength,
(corresponding to the idler mode) $ |E_i\rangle+|L_i\rangle $ and a hybrid
light-solid qubit $ |E_{QM}\rangle+|L_s\rangle $
that is a superposition between a single-atomic excitation delocalized in a 
solid and a photonic state (associated to the signal mode).

Instead of using the double read-out to analyze the signal photon, only the
early mode $|E_s\rangle$ is stored in the AFC. It is released after a time
delay of exactly $\tau$. The detection of this mode, which we will hereafter
refer to as \emph{echo}, is thus made indistinguishable from the detection of the late
$|L_s\rangle$ mode, which is directly transmitted through the memory. Taking the
echo efficiency $\eta_{\rm{echo}}$ and the transmission probability $\eta_{\rm{trans}}$ into account, this corresponds to a projection onto the vector
\begin{equation}
  \cos{\theta} \langle L_s| +e^{i\phi_s}  \sin{\theta} \langle E_s|,
\end{equation}
with
$
\cos{\theta}=\sqrt{\frac{\eta_{\rm{trans}}}{\eta_{\rm{trans}}+\eta_{\rm{echo}}}},
$
and
$
\sin{\theta}=\sqrt{\frac{\eta_{\rm{echo}}}{\eta_{\rm{trans}}+\eta_{\rm{echo}}}}.
$
The phase factor is controlled via the AFC structure~\cite{Afzelius2009a} and
is chosen to be either $\phi_{s,1}=0^\circ$ or $\phi_{s,2}=180^\circ.$ This 
corresponds to measuring the following operators with eigenvalues $\{+1, -1\}$,
\begin{eqnarray}
  && \nonumber X_1=\sin{2\theta} \-\ \sigma_x +\cos{2\theta} \-\ \sigma_z, \\
  && \nonumber X_2=-\sin{2\theta} \-\ \sigma_x +\cos{2\theta} \-\ \sigma_z,
\end{eqnarray}
where $\sigma_x$ and $\sigma_z$ are Pauli matrices.

Fifty metres away, the idler photon, at telecommunication wavelength, is
projected either onto the $z$-direction (corresponding to the operator $Y_1=\sigma_z$) by
measuring the time of arrival at the detector or onto the $x$-direction
($Y_2=\sigma_x$) using a Michelson interferometer. One then finds for the
correlators,
\begin{eqnarray}
  \nonumber && E(X_1 Y_1)=E(X_2 Y_1)=\cos(2\theta)\\
  \nonumber && E(X_1 Y_2)=-E(X_2 Y_2)=\sin(2\theta).
\end{eqnarray}\\
so that the CHSH polynomial \mbox{$S=2\cos(2\theta)+2\sin(2\theta)$} is
maximized to $2\sqrt{2}$ for 
$\cos{2\theta}=\frac{\eta_{\rm{trans}}-\eta_{\rm{echo}}}{\eta_{\rm{trans}}+\eta_{\rm{echo}}}=\frac{\sqrt{2}}{2},
$
i.e.~for a ratio between the echo and transmitted pulses
\begin{equation}
  \label{cond_optviolation}
  \frac{\eta_{\rm{echo}}}{\eta_{\rm{trans}}} \approx \frac{1}{5.8},
\end{equation}
which corresponds to the bases indicated in Fig.~\ref{fig:bloch-sphere}b. This
result is intuitive, as it simply corresponds to the settings of
Fig.~\ref{fig:bloch-sphere}a rotated around the $x$-axis.

In the experiment, the AFC structure is modified to satisfy this requirement. (We
measured $\eta_{\rm{trans}} \approx 0.36,$ $\eta_{\rm{echo}} \approx 0.05$,
giving a ratio of $1/7.2$). Under the assumption that the marginals are
the same for the telecom photon, independent of the result of the measurement on the signal mode, we
measure
\begin{eqnarray}
  \nonumber && E(X_1 Y_1)=0.68\pm0.05\\
  \nonumber && E(X_2 Y_1)=0.71\pm 0.06\\
  \nonumber && E(X_1 Y_2)=0.63 \pm 0.09\\
  \nonumber && E(X_2 Y_2)=-0.60\pm0.09
\end{eqnarray}
leading to $S=2.62 \pm 0.15$, a violation of
the CHSH inequality by more than 4 standard deviations. This clearly shows that
when the early mode was stored in our solid, the two qubit state involving the hybrid qubit and the
telecom qubit was entangled.

Taking a closer look, it might at first be surprising that the violation of the
Bell inequality can be maximal, since directly after the absorption the system is
described by the \emph{non}-maximally entangled state (neglecting noise)
\begin{equation}
  \label{nonmax_ent}
  \frac{1}{\sqrt{1+\alpha^2}}\left(\alpha|E_{QM}E_i\rangle+|L_{s}L_{i}\rangle\right),
\end{equation}
where $\alpha$ is related to the absorption efficiency $\eta_{\rm abs}$
of the quantum memory by $\alpha = \sqrt{\eta_{\rm{abs}}}$. (The absorption
efficiency corresponds to the absorption
by the comb peaks and does not include the absorption by residual atoms whose
resonance frequencies fall between the peaks, so that $\eta_{\rm{trans}} \neq
1-\eta_{\rm abs}$.) The explanation is that the memory based
measurement is a generalized measurement. More precisely, assuming that
$\eta_\text{echo} = \eta_\text{abs}^2 \eta$~\cite{Afzelius2009a}, the
measurement consists of two generalized measurements $\{X_1',X_2'\}$ made 
from $\{\Pi_{+1}^s, \Pi_{-1}^s, 1-\Pi_{+1}^s-\Pi_{-1}^s\},$  i.e.~projections onto
the non-orthogonal vectors
\begin{eqnarray}
  \nonumber && \Pi_{+1}^s : \sqrt{\eta_{\rm{trans}}} \langle L_s| + e^{i\phi_s}
  \sqrt{\eta \eta_{\rm{abs}}} \langle E_{QM}|,\\ \nonumber &&  \Pi_{-1}^s :
  e^{i(\phi_s+\pi)} \sqrt{\eta} \eta_{\rm{abs}} \langle L_s| +
  \sqrt{\frac{\eta_{\rm{trans}}}{\eta_{\rm{abs}}}} \langle E_{QM}|,
\end{eqnarray}
where $\phi_s$ is controlled from the AFC structure to be  either
$\phi_{s,1}=0^\circ$ or $\phi_{s,2}=180^\circ$ for $X_1$ and $X_2$ respectively.
Under the fair sampling assumption where inconclusive results are discarded, one can only take
successful projections on  $\Pi_{+1}^s$ and $\Pi_{-1}^s$ into account. If one
assigns the value $+1$ ($-1$) to a conclusive projection into $\Pi_{+1}^s$
($\Pi_{-1}^s$), one finds random marginals and a Bell violation of $2\sqrt{2}$
provided that the condition (\ref{cond_optviolation}) is fulfilled. This is
analogue to the distillation of entanglement reported in ref.~\cite{Kwiat2001}
where a non unitary filtering process equalizes the contributions of the two
terms in Eq.~(\ref{nonmax_ent}), thereby yielding optimal CHSH violation from a
non-maximally entangled state.

\bibliography{QuantumStorage}
\end{document}